\pdfoutput=1

\documentclass[11pt]{article}

\usepackage[]{EMNLP2022}

\usepackage{times}
\usepackage{latexsym}

\usepackage[T1]{fontenc}

\usepackage[utf8]{inputenc}

\usepackage{microtype}

\usepackage{inconsolata}

\usepackage{amsfonts}
\usepackage{amsmath}
\usepackage{amssymb}
\usepackage{algorithm,algorithmicx,algpseudocode}
\usepackage{bm}
\usepackage{booktabs}
\usepackage{cancel}
\usepackage{multirow}
\usepackage{graphicx}
\usepackage{arydshln}
\usepackage{hyperref}
\usepackage{mathtools}
\usepackage{xcolor,soul}
\sethlcolor{lightgray}

\definecolor{red}{rgb}{1,0,0}
\newcommand{\red}[1]{{\color{red} #1}}
\definecolor{blue}{rgb}{0,0,1}
\newcommand{\blue}[1]{{\color{blue} #1}}

\DeclareMathOperator*{\argmax}{\mathrm{argmax}}

\makeatletter
\def\adl@drawiv#1#2#3{%
        \hskip.5\tabcolsep
        \xleaders#3{#2.5\@tempdimb #1{1}#2.5\@tempdimb}%
                #2\z@ plus1fil minus1fil\relax
        \hskip.5\tabcolsep}
\newcommand{\cdashlinelr}[1]{%
  \noalign{\vskip\aboverulesep
           \global\let\@dashdrawstore\adl@draw
           \global\let\adl@draw\adl@drawiv}
  \cdashline{#1}
  \noalign{\global\let\adl@draw\@dashdrawstore
           \vskip\belowrulesep}}
\makeatother
\newcommand{\pz}{\phantom{0}}

\usepackage{cleveref}
\crefformat{section}{\S#2#1#3}
\crefformat{subsection}{\S#2#1#3}
\crefformat{subsubsection}{\S#2#1#3}

\usepackage{tikz}
\usepackage{pgfplots}
\pgfplotsset{width=7.5cm,compat=1.12}
\usepgfplotslibrary{fillbetween}

\title{BERT Meets CTC: New Formulation of End-to-End Speech Recognition\\with Pre-trained Masked Language Model}

\author{
    Yosuke Higuchi$^{1,2}$, Brian Yan$^{1}$, Siddhant Arora$^1$, Tetsuji Ogawa$^2$,\\
    {\bf Tetsunori Kobayashi}$^2$, {\bf Shinji Watanabe}$^1$\\
    $^1$Carnegie Mellon Univeristy, $^2$Waseda University\\
    \texttt{higuchi@pcl.cs.waseda.ac.jp} \\
}

\begin{document}
\maketitle

\begin{abstract}
This paper presents BERT-CTC,
a novel formulation of end-to-end speech recognition
that adapts BERT for connectionist temporal classification (CTC).
Our formulation relaxes the conditional independence assumptions used in conventional CTC and incorporates linguistic knowledge through the explicit output dependency obtained by BERT contextual embedding.
BERT-CTC attends to the full contexts of the input and hypothesized output sequences via the self-attention mechanism.
This mechanism encourages a model to learn inner/inter-dependencies between the audio and token representations while maintaining CTC's training efficiency.
During inference,
BERT-CTC combines a mask-predict algorithm with CTC decoding,
which iteratively refines an output sequence.
The experimental results reveal that BERT-CTC improves over conventional approaches
across variations in speaking styles and languages.
Finally, we show that the semantic representations in BERT-CTC are beneficial towards downstream spoken language understanding tasks.
\end{abstract}

\section{Introduction}
\label{sec:intro}
The field of natural language processing (NLP) has witnessed remarkable improvements in performance
thanks to the advances in deep learning-based techniques~\cite{collobert2011natural,bahdanau2014neural,sutskever2014sequence,vaswani2017attention,young2018recent}.
Much of the recent progress in NLP lies in large-scale language models (LMs)~\cite{devlin2019bert,brown2020language},
which are pre-trained on a vast amount of text data to learn versatile linguistic knowledge~\cite{tenney2019bert}.
Such pre-trained models have been shown to improve diverse NLP tasks,
alleviating the heavy requirement of supervised training data.
Inspired by the great success in NLP,
pre-trained LMs have been actively adopted for speech processing tasks, including
automatic speech recognition (ASR)~\cite{shin2019effective,huang2021speech},
spoken language understanding (SLU)~\cite{chuang2020speechbert,chung2021splat}, and
text-to-speech synthesis~\cite{hayashi2019pre,kenter2020improving}.

This paper focuses on leveraging pre-trained LMs for end-to-end ASR (E2E-ASR),
which aims to model direct speech-to-text conversion~\cite{graves2014towards,chorowski2015attention,chan2016listen}.
One of the challenges in E2E-ASR is a huge discrepancy between input and output sequences;
the input is a continuous acoustic signal with fine-grained patterns,
while the output is discrete linguistic symbols (e.g., words)
with long-range dependencies.
Such an input-output gap makes it difficult for an E2E-ASR model to
extract semantic/morphosyntax information from speech,
which is essential for generating proper text.
We believe this limitation can be mitigated by
taking advantage of the rich linguistic representations obtained from pre-trained LMs.

Several attempts have been made to use pre-trained LMs indirectly for improving E2E-ASR,
such as N-best hypothesis rescoring~\cite{shin2019effective,salazar2020masked,chiu2021innovative,futami2021asr,udagawa2022effect} and
knowledge distillation~\cite{futami2020distilling,bai2021fast,kubo2022knowledge}.
Others have investigated directly unifying an E2E-ASR model with a pre-trained LM,
where the LM is fine-tuned to optimize ASR in an end-to-end trainable framework~\cite{huang2021speech,yi2021efficiently,zheng2021wav,deng2021improving,yu2022non}.

We explore a novel direction for adopting a pre-trained masked language model (MLM) for E2E-ASR,
based on connectionist temporal classification (CTC)~\cite{graves2006connectionist}.
Compared to other autoregressive approaches, such as RNN-Transducer (RNN-T)~\cite{graves2012sequence} and attention-based sequence-to-sequence~\cite{chorowski2015attention},
CTC's non-autoregressive formulation allows simple training and inference processes for realizing E2E-ASR.
However, the performance of CTC is often limited due to a \textit{conditional independence assumption}
between output tokens~\cite{chiu2018state}.
In this work, we propose \textbf{BERT-CTC} that adapts BERT~\cite{devlin2019bert} for CTC to
mitigate the conditional independence assumption.
BERT-CTC conditions CTC outputs on context-aware BERT embeddings,
thereby incorporating explicit linguistic information into training/inference.
The BERT-conditional formulation enables a model to attend to the full contexts of the input and hypothesized output sequences via the self-attention mechanism,
while maintaining the benefits of a simple training algorithm in CTC.
During inference,
BERT-CTC combines a mask-predict algorithm with CTC decoding,
which iteratively refines outputs with flexible length adjustment.

The key contributions of this work are summarized as follows:
\begin{itemize}
    \vspace{-0.15cm}
    \setlength{\itemsep}{0cm}
    \item We propose BERT-CTC, which efficiently adapts a pre-trained MLM for CTC-based E2E-ASR without fine-tuning. We provide a probabilistic formulation of our BERT-CTC and its close relation to conventional approaches, i.e., CTC and RNN-T.
    \item We evaluate BERT-CTC in various ASR tasks, which demonstrates its effectiveness regardless of variations in speaking styles and languages. We also show its potential application to end-to-end SLU.
    \item The codes and recipes are open-sourced on ESPnet~\cite{watanabe2018espnet}, the widely used toolkit for end-to-end speech processing.\footnote{\url{https://github.com/YosukeHiguchi/espnet/tree/bert-ctc}} We hope our work encourages further research on combining ASR with pre-trained LMs, helping to bridge ASR and NLP fields.
\end{itemize}

\section{Background}
To understand how BERT-CTC exploits BERT for
relaxing the conditional independence assumption in CTC,
we start with a brief review of probabilistic formulations of conventional E2E-ASR approaches,
including CTC~\cite{graves2006connectionist,graves2014towards} and RNN-T~\cite{graves2012sequence,graves2013speech}.

\paragraph{Definition of End-to-End ASR} Let $O=(\bm{\mathrm{o}}_t\in\mathbb{R}^D| t=1,\cdots,T)$ be an input sequence of length $T$, and
$W =( w_n\in\mathcal{V} | n=1,\cdots,N )$ be the corresponding output sequence of length $N$.
Here, $\bm{\mathrm{o}}_t$ is a $D$-dimensional acoustic feature at frame $t$, 
$w_n$ is an output token at position $n$, and $\mathcal{V}$ is a vocabulary.\footnote{We consider $\mathcal{V}$ as a vocabulary constructed for pre-training a large-scale MLM, i.e., BERT.}
In general,
the output length is much shorter than the input length (i.e., $N\!\ll \!T$).
The objective of ASR is to find the most probable output sequence $\hat{W}$
that corresponds to a given input sequence $O$:
\begin{equation}
    \hat{W} = \argmax_{W \in \mathcal{V}^{*}} p(W|O), \label{eq:asr}
\end{equation}
where $\mathcal{V}^{*}$ denotes all possible token sequences.
E2E-ASR aims to realize the direct mapping from $O$ to $W$ by
modeling the posterior distribution $p(W|O)$ with a single deep neural network.

\subsection{Connectionist Temporal Classification}
\label{ssec:ctc}
CTC formulates E2E-ASR by
considering all possible alignments between an input sequence $O$ and output sequence $W$.
To align the sequences at the frame level,
CTC augments an output sequence by allowing repetitions of the same token and
inserting a blank symbol $\epsilon$ for representing ``no output token'' (e.g., silence).
Let $A$ denote an augmented output sequence defined as
$A=(a_t\in\mathcal{V} \cup \{\epsilon\} | t=1,\cdots,T)$, 
which we refer to as an \textit{alignment} between $O$ and $W$.

With the introduction of the frame-level alignment,
CTC factorizes $p(W|O)$ as follows:
\begin{align}
    p_{\mathsf{ctc}}(W|O) &= \sum_{A \in \mathcal{B}_{\mathsf{ctc}}^{-1}(W)} p(W|A,\cancel{O}) p(A|O) \label{eq:p_ctc_W_O} \\
    &\approx \sum_{A \in \mathcal{B}_{\mathsf{ctc}}^{-1}(W)} p(A|O), \label{eq:p_ctc_W_O_approx}
\end{align}
where $\mathcal{B}_{\mathsf{ctc}}$ is the collapsing function~\cite{graves2006connectionist} that maps $A$ to $W$ by suppressing repeated tokens and removing blank symbols, and
$\mathcal{B}_{\mathsf{ctc}}^{-1}(W)$ is a set of all possible CTC alignments that are compatible with $W$.
To obtain Eq.~\eqref{eq:p_ctc_W_O_approx}, 
CTC makes a conditional independence assumption of $O$ in Eq.~\eqref{eq:p_ctc_W_O}, and
we assume $p(W|A)=1$, as $W$ can be determined uniquely by the collapsing function.

The joint probability $p(A|O)$ is further factorized using the probabilistic chain rule as
\begin{equation}
    p(A|O) \approx \prod_{t=1}^{T} p(a_t|\cancel{a_1,\cdots,a_{t-1}},O). \label{eq:p_ctc_A_O_approx}
\end{equation}
In Eq.~\eqref{eq:p_ctc_A_O_approx},
CTC makes a conditional independence assumption between output tokens, where
$p(A|O)$ is approximated as the product of token emission probabilities at each time frame.
The conditional probability $p(a_t|O)$ in Eq.~\eqref{eq:p_ctc_A_O_approx} is computed as
\begin{align}
    p(a_t|O) &= \text{Softmax}(\text{Linear}(\bm{\mathrm{h}}^{\mathsf{ae}}_t)), \label{eq:p_ctc_a_o} \\
    \bm{\mathrm{h}}^{\mathsf{ae}}_t &\sim \text{AudioEnc} (O). \label{eq:h_t}
\end{align}
In Eq.~\eqref{eq:p_ctc_a_o},
$\text{Softmax}(\cdot)$ is a softmax function, and
$\text{Linear}(\cdot)$ is a linear projection layer.
$\text{AudioEnc}(\cdot)$ in Eq.~\eqref{eq:h_t} is an audio encoder network that embeds speech input into a sequence of $d^{\mathsf{ae}}$-dimensional hidden vectors $H^{\mathsf{ae}}=(\bm{\mathrm{h}}^{\mathsf{ae}}_t\in\mathbb{R}^{d^{\mathsf{ae}}}|t=1,\cdots,T)$.

\paragraph{Training}
The objective function of CTC is defined by the negative log-likelihood of Eq.~\eqref{eq:p_ctc_A_O_approx} over all possible alignments:
\begin{align}
    \mathcal{L}_{\mathsf{ctc}} (O, W) = -\log \sum_{A \in \mathcal{B}_{\mathsf{ctc}}^{-1}(W)} \prod_{t=1}^{T} p(a_t|O).
    \label{eq:L_ctc}
\end{align}
The summation in Eq.~\eqref{eq:L_ctc} is efficiently computed via dynamic programming~\cite{graves2006connectionist}.

\paragraph{Inference}
Eq.~\eqref{eq:asr} is solved using the best path decoding algorithm~\cite{graves2006connectionist}.
The algorithm first obtains the most probable alignment $\hat{A}$ in a greedy manner,
concatenating the most active tokens at each frame: $\hat{a}_t=\argmax_{a_t} p(a_t|O)$.
The most probable token sequence $\hat{W}$ is then obtained by applying the collapsing function to $\hat{A}$ as $\hat{W} = \mathcal{B}_{\mathsf{ctc}}(\hat{A})$.

\subsection{RNN-Transducer}
\label{ssec:rnnt}
CTC estimates
the distribution over alignments only depending on speech input (Eq.~\eqref{eq:p_ctc_A_O_approx}).
Thus, by definition, CTC cannot consider output dependencies,
preventing a model from properly capturing the multimodal distribution of target token sequences~\cite{gu2018non}.

RNN-T overcomes this problem by making each token prediction explicitly conditioned on the previous non-blank output tokens $(w_1,\cdots,w_{n-1})$.
Let $Z=(z_u\in\mathcal{V}\cup\{\epsilon\}|u=1,\cdots,T+N)$ be an alignment
used in RNN-T, and
RNN-T factorizes $p(W|O)$ similarly to Eq.~\eqref{eq:p_ctc_W_O_approx} as
\begin{equation}
    p_{\mathsf{rnnt}}(W|O) \approx \sum_{Z\in\mathcal{B}_{\mathsf{rnnt}}^{-1}(W)} p(Z|O), \label{eq:p_rnnt_W_O_approx}
\end{equation}
where $\mathcal{B}_{\mathsf{rnnt}}$ is the collapsing function of RNN-T~\cite{graves2012sequence}
that map $Z$ to $W$.
The joint probability $p(Z|O)$ is factorized using the probabilistic chain rule \textit{without} the conditional independence assumption (cf. Eq.~\eqref{eq:p_ctc_A_O_approx}) as
\begin{align}
    p(Z|O) &= \prod_{u=1}^{T+N} p(z_{u}|z_1,\cdots,z_{u-1},O) \label{eq:p_rnnt_A_O} \\
    &\approx \prod_{u=1}^{T+N} p(z_{u}|\underbrace{w_1,\cdots,w_{n_u-1}}_{= \mathcal{B}_{\mathsf{rnnt}}(z_1,\cdots,z_{u-1})},O), \label{eq:p_rnnt_A_O_approx} 
\end{align}
where $n_u$ is the number of tokens predicted up to an index of $u$.
From Eq~\eqref{eq:p_rnnt_A_O} to Eq.~\eqref{eq:p_rnnt_A_O_approx}, RNN-T assumes $(z_1,\cdots,z_{u-1})\approx(w_1,\cdots,w_{n_u-1})$,
which is reasonable since $W$ can be determined uniquely by the collapsing function.
The conditional probability $p(z_u|w_1,\cdots,w_{n_u-1},O)$ is computed as
\begin{align}
    p(z_{u}|&w_1,\cdots,w_{n_u-1},O) \nonumber \\ 
    &= \text{Softmax}(\text{JointNet}(\bm{\mathrm{h}}^{\mathsf{ae}}_t, \bm{\mathrm{h}}^{\mathsf{pn}}_{n_u})), \label{eq:p_rnnt_a_w_O}\\
    \bm{\mathrm{h}}^{\mathsf{pn}}_{n_u} &= \text{PredictionNet} (w_1,\cdots,w_{n_u-1}). \label{eq:h_n}
\end{align}
In Eq.~\eqref{eq:p_rnnt_a_w_O},
$\bm{\mathrm{h}}^{\mathsf{ae}}_t$ is obtained from the audio encoder (Eq.~\eqref{eq:h_t}),
and $\text{JointNet}(\cdot)$ is a joint network that combines the audio and token representations,
$\bm{\mathrm{h}}^{\mathsf{ae}}_t$ and $\bm{\mathrm{h}}^{\mathsf{pn}}_{n_u}$,
using a linear projection layer.
In Eq.~\eqref{eq:h_n},
$\text{PredictionNet}(\cdot)$ is a prediction network that
encodes the previous non-blank output tokens to a hidden vector $\bm{\mathrm{h}}_{n_u}^{\mathsf{pn}}$.
The adoption of the prediction network is the main difference from CTC,
which explicitly captures causal dependency in outputs.

\paragraph{Training}
The RNN-T loss $\mathcal{L}_{\mathsf{rnnt}} (O, W)$ is defined by the negative log-likelihood of Eq.~\eqref{eq:p_rnnt_A_O_approx}.
Similar to the CTC objective in Eq.~\eqref{eq:L_ctc}, the summation over alignments is efficiently computed using dynamic programming~\cite{graves2012sequence}.

\paragraph{Inference}
RNN-T estimates the most probable token sequence $\hat{W}$ using the beam search algorithm proposed in~\cite{graves2012sequence}.

\begin{figure*}[t]
    \begin{minipage}[b]{0.23\linewidth}
        \centering
        \includegraphics[height=5.3cm]{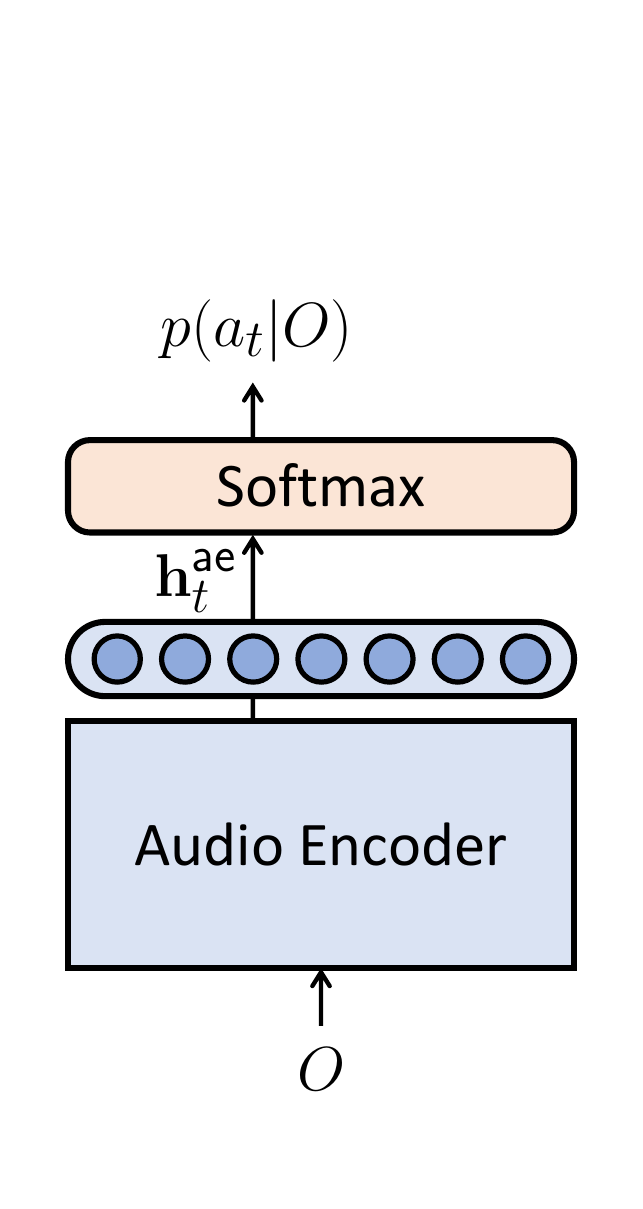}\vspace{-0.4cm}
        \hspace{0.65cm} (a) CTC
    \end{minipage}
    \begin{minipage}[b]{0.38\linewidth}
        \centering
        \hspace{-0.34cm}
        \includegraphics[height=5.3cm]{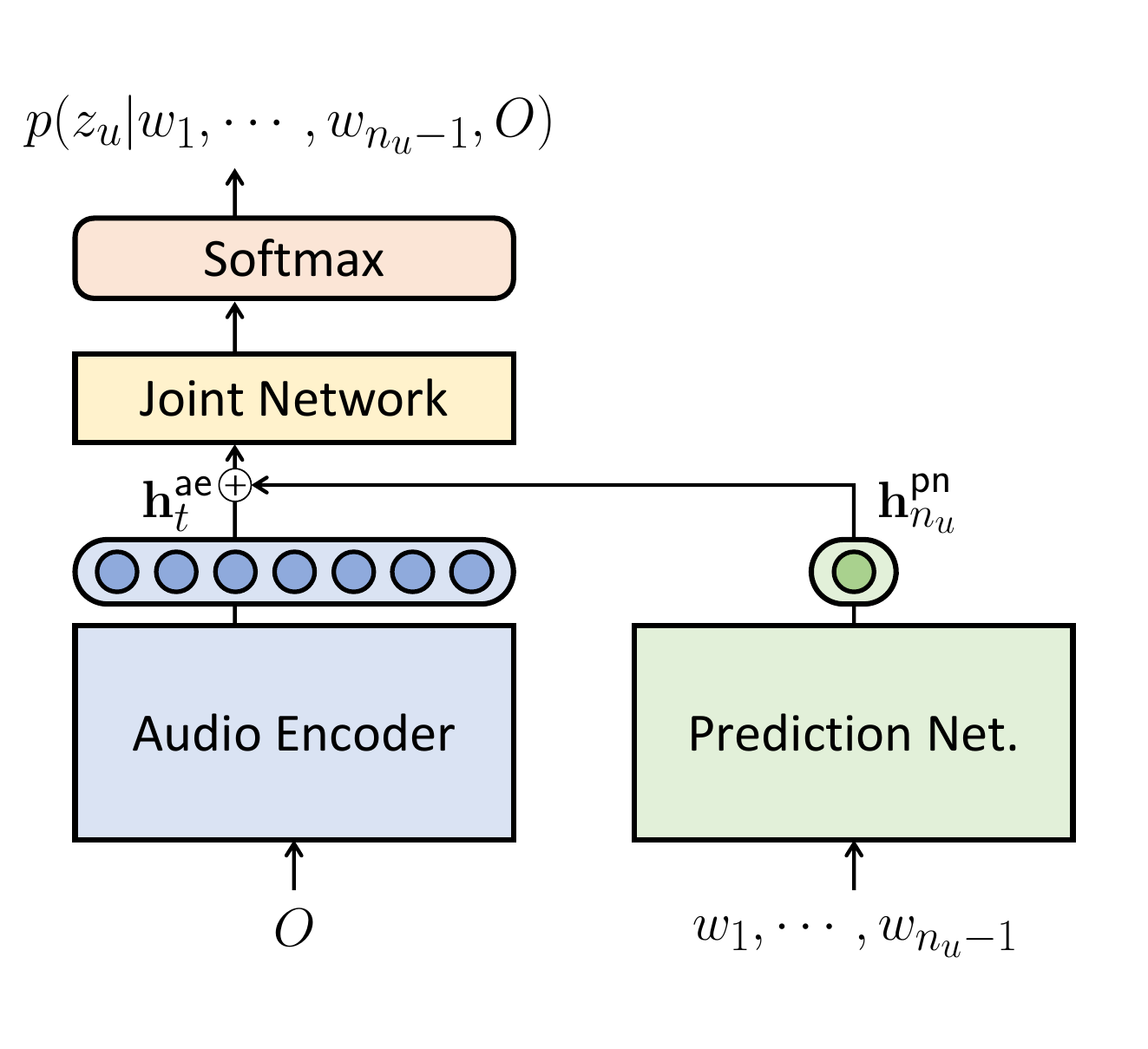}\vspace{-0.4cm}
        \hspace{0.65cm} (b) RNN-T \hspace{0.1cm}
    \end{minipage}
        \begin{minipage}[b]{0.38\linewidth}
        \centering
        \includegraphics[height=5.3cm]{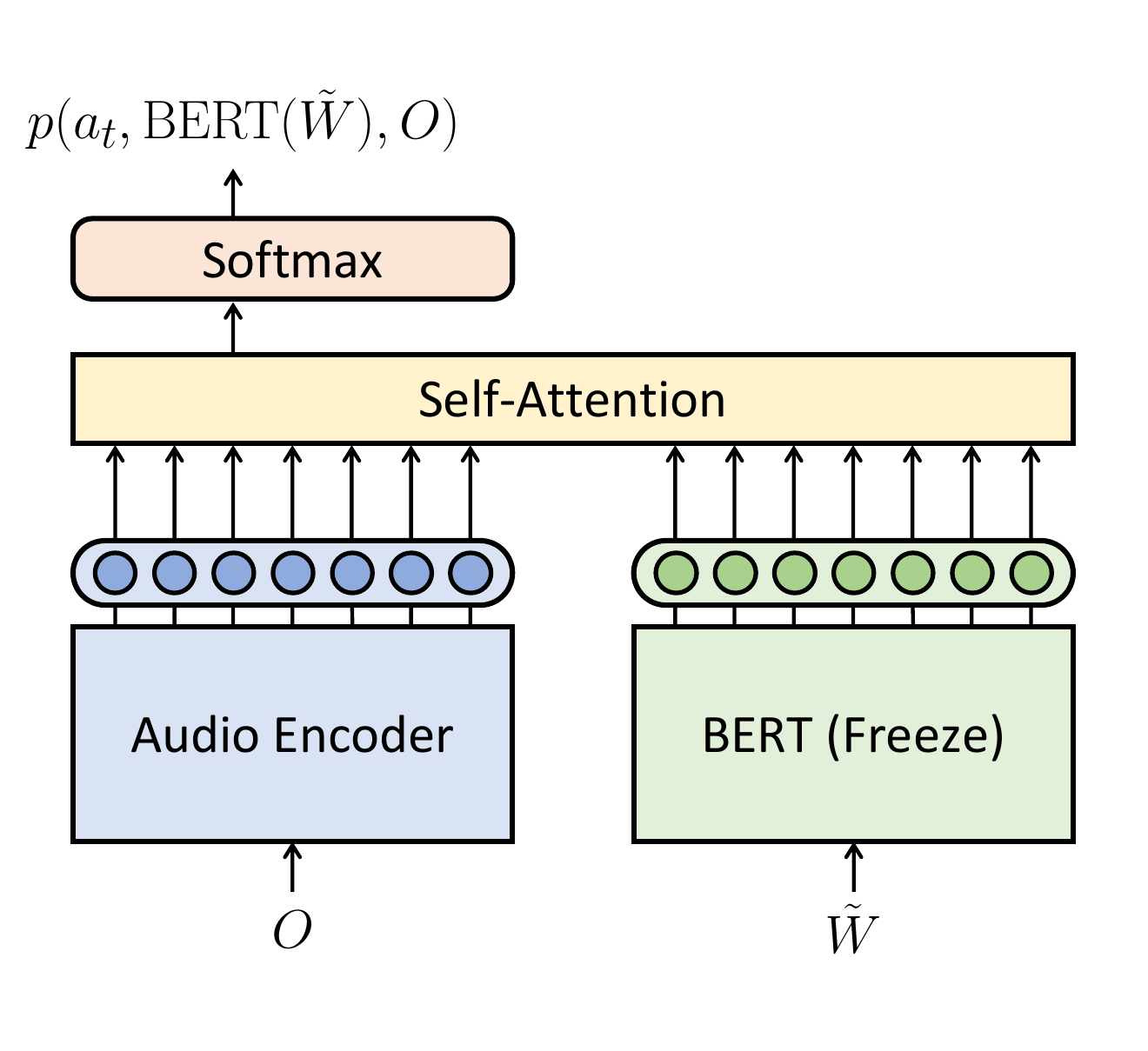}\vspace{-0.4cm}
        \hspace{0.65cm} (c) BERT-CTC (ours)
    \end{minipage}
    \vspace{-0.3cm}
    \caption{Comparisons between different model architectures for end-to-end ASR.}
    \label{fig:e2easr}
\end{figure*}

\section{BERT-CTC}

\paragraph{Overview}
In Fig.~\ref{fig:e2easr},
we compare our proposed E2E-ASR model, \textbf{BERT-CTC}, to CTC and RNN-T.
BERT-CTC leverages powerful representations from BERT~\cite{devlin2019bert}
to make CTC training/inference explicitly conditioned on linguistic information (Fig.~\ref{fig:e2easr}(a) vs.\ Fig.~\ref{fig:e2easr}(c)).
We use BERT as a feature extractor for a (masked) token sequence,
whose parameters are frozen during training.
BERT-CTC can be similar to RNN-T in that
audio and token representations are fused to estimate
the distribution over alignments (Fig.~\ref{fig:e2easr}(b) vs.\ Fig.~\ref{fig:e2easr}(c)).
However, BERT-CTC attends to the full contexts of the input and output sequences via the self-attention mechanism~\cite{vaswani2017attention},
which facilitates a model to learn inner/inter-dependencies within/between the sequences.

BERT-CTC is formulated by introducing a partially masked (or partially observed) sequence
$\tilde{W}=(\tilde{w}_n\in\mathcal{V} \cup \{\texttt{[MASK]}\} | n=1,\cdots,N)$,
which is obtained by replacing some tokens in an output sequence $W$ with a special mask token $\texttt{[MASK]}$.
Note that during inference,
we apply masks to a hypothesized sequence $\hat{W}$ to obtain a masked sequence.
Considering all possible $\tilde{W}$,
the conditional probability $p(W|O)$ is factorized as follows:
\begin{align}
    \hspace{-0.08cm}
    p_{\mathsf{bc}}&(W|O)
    =\!\sum_{\tilde{W}\!\in \mathcal{A}(W)} p(W|\tilde{W},O) p(\tilde{W}|O), \label{eq:p_bertctc_W_O}
\end{align}
where $\mathcal{A}(W)$ covers $W$ with all possible masking patterns.
Here, we interpret $p(\tilde{W}|O)$ as
a prior distribution of sequences consisting of observed tokens that are easily recognized only from speech input;
the other masked tokens are difficult and require contextual information to be determined (e.g., homophones),
which is modeled by $p(W|\tilde{W},O)$.
We further describe the above interpretation in the training (\cref{ssec:bertctc_training}) and inference (\cref{ssec:bertctc_inference}) sections.

The conditional probability $p(W|\tilde{W},O)$ is further factorized by using the CTC alignment as
\begin{align}
    &p(W|\tilde{W},O) = \sum_{A \in \mathcal{B}_{\mathsf{ctc}}^{-1}(W)} p(W,A|\tilde{W},O) \\
    &\approx \sum_{A \in \mathcal{B}_{\mathsf{ctc}}^{-1}(W)} p(A|W,\cancel{\tilde{W}},O) p(W|\tilde{W},\cancel{O}). \label{eq:p_W_tW_O_approx}
\end{align}
In Eq.~\eqref{eq:p_W_tW_O_approx},
we make two conditional independence assumptions.
The first is that given $W$ and $O$,
$\tilde{W}$ is not required to determine $A$.
This is reasonable because $W$ already contains observed tokens in $\tilde{W}$ and
is helpful in avoiding the combination of all possible masked sequences and alignments (i.e., $\mathcal{A}\times\mathcal{B}_{\mathsf{ctc}}^{-1}$).
The second is that given $\tilde{W}$, $O$ is not required to determine $W$.
We consider $p(W|\tilde{W})$ as a strong prior modeled by a pre-trained MLM (i.e., BERT),
which can be achieved without the observation from $O$.
We empirically show that this assumption holds in~\cref{ssec:exp_cia_O}.

Similar to CTC, the joint probability $p(A|W,O)$ is factorized using the probabilistic chain rule as
\begin{align}
    \hspace{-0.15cm}
    p(A|W,O)
    &\approx \prod_{t=1}^{T} p(a_t|\cancel{a_1,\cdots,a_{t-1}},W,O). \label{eq:p_bertctc_A_W_O_approx}
\end{align}
To obtain Eq.~\eqref{eq:p_bertctc_A_W_O_approx},
we make the same conditional independence assumption as in CTC.
However, compared to Eq.~\eqref{eq:p_ctc_A_O_approx},
Eq.~\eqref{eq:p_bertctc_A_W_O_approx} is conditioned on an output sequence $W$,
enabling a model to explicitly use linguistic information to estimate the distribution over alignments.
This is somewhat similar to RNN-T (Eq.~\eqref{eq:p_rnnt_A_O_approx}),
but is different in that BERT-CTC attends to the whole context $(w_1,\cdots,w_N)$.
We discuss this advantage in~\cref{ssec:exp_ablation}.

Substituting Eq.~\eqref{eq:p_bertctc_A_W_O_approx} into Eq.~\eqref{eq:p_W_tW_O_approx},
we model the product of $p(a_t|W,O)$ and $p(W|\tilde{W})$ as
\begin{equation}
    \text{Eq.~\eqref{eq:p_W_tW_O_approx}} \triangleq \sum_{A \in \mathcal{B}_{\mathsf{ctc}}^{-1}(W)} \prod_{t=1}^{T} p(a_t|\text{BERT}(\tilde{W}),O), \label{eq:p_W_tW_O_approx_bert}
\end{equation}
where $\text{BERT}(\cdot)$ is the output of BERT representing the distribution of target sequences.\footnote{Note that $\text{BERT}(\cdot)$ can be any pre-trained MLM.}
This enables Eq.~\eqref{eq:p_W_tW_O_approx_bert} to be realized with a single differentiable model,
enabling the whole network to be trained end-to-end.
The conditional probability $p(a_t|\text{BERT}(\tilde{W}),O)$ is computed as
\begin{align}
    p(a_t|&\text{BERT}(\tilde{W}),O) \nonumber \\
    &= \text{Softmax}(\text{SelfAttn}_t(H^{\mathsf{ae}}, H^{\mathsf{bert}})), \label{eq:p_bertctc_a_tW_O} \\
    H^{\mathsf{bert}} &= \text{BERT}(\tilde{W}). \label{eq:H_bert}
\end{align}
In Eq.~\eqref{eq:p_bertctc_a_tW_O},
$\text{SelfAttn}_t(\cdot)$ indicates the $t$-th output of stacked Transformer self-attention layers~\cite{vaswani2017attention},
which consume the concatenated $H^{\mathsf{ae}}$ (from Eq.~\eqref{eq:h_t}) and $H^{\mathsf{bert}}$.\footnote{We apply simple embedding layers to $H^{\mathsf{ae}}$ and $H^{\mathsf{bert}}$ so that the dimensions of hidden vectors match, but we omit it for simplicity. See Appendix~\ref{apdx:model_conf} for detailed implementation.}
In Eq.~\eqref{eq:H_bert},
$\text{BERT}(\cdot)$ embeds a masked sequence $\tilde{W}$ into a sequence of $d^{\mathsf{bert}}$-dimensional hidden vectors $H^{\mathsf{bert}}= (\bm{\mathrm{h}}^{\mathsf{bert}}_n\in\mathbb{R}^{d^{\mathsf{bert}}}|n=1,\cdots,N)$.

\subsection{Training}\label{ssec:bertctc_training}
The BERT-CTC objective is defined by the negative log-likelihood of Eq.~\eqref{eq:p_bertctc_W_O} expanded with Eq.~\eqref{eq:p_W_tW_O_approx}:
\begin{align}
    \hspace{-0.07cm}
    -\log \sum_{\tilde{W}} \sum_{A} p(A|W,O) p(W|\tilde{W})p(\tilde{W}|O).
    \label{eq:L_bertctc}
\end{align}
To deal with the intractable marginalization over $\tilde{W}$ in Eq.~\eqref{eq:L_bertctc},
we rewrite it under expectation with respect to the probability distribution $p(\tilde{W}|O)$:
\begin{equation}
    \approx -\log \mathbb{E}_{\tilde{W} \sim p(\tilde{W}|O)} \bigg[\sum_{A} p(A|W,O) p(W|\tilde{W})\bigg], \nonumber
\end{equation}
whose upper bound can be derived by using the Jensen's inequality as
\begin{align}
    &\hspace{-0.29cm}\le -\mathbb{E}_{\tilde{W}} \bigg[ \log \sum_{A} p(A|W,O) p(W|\tilde{W})\bigg] \nonumber \\
    &\hspace{-0.29cm}\approx \underbrace{-\mathbb{E}_{\tilde{W}} \bigg[ \log \sum_{A} \prod_{t} p(a_t|\text{BERT}(\tilde{W}),O) \bigg] }_{{\triangleq \mathcal{L}_{\mathsf{bc}}(O,W)}}, \label{eq:L_bertctc_approx}
\end{align}
where $\mathcal{L}_{\mathsf{bc}}$ is the loss for BERT-CTC training.

Compared with the CTC objective (Eq.~\eqref{eq:L_ctc}),
each token prediction in Eq.~\eqref{eq:L_bertctc_approx} is explicitly conditioned on contextual embedding from BERT.
This relaxes the conditional independence assumption between outputs
while retaining the same optimization strategy as in CTC.
For sampling $\tilde{W}$ in Eq.~\eqref{eq:L_bertctc_approx},
we use random sampling from a uniform distribution to approximate the probability distribution of $p(\tilde{W}|O)$,
for the sake of simplicity.
We first obtain the random number of tokens from a uniform distribution as $M\sim\text{Uniform}(1, N)$.
Then, $M$ tokens in a ground-truth sequence $W$ are randomly selected to be replaced with \texttt{[MASK]}, similar to~\cite{ghazvininejad2019mask}.

\paragraph{Hierarchical Loss}
We apply an auxiliary CTC loss to the audio encoder output in a hierarchical multi-tasking manner~\cite{fernandez2007sequence,sanabria2018hierarchical}.
As the vocabulary size of BERT is often too large for ASR training,
we train the audio encoder to predict a sequence $W'=(w'_l\in\mathcal{V}'|l=1,\cdots,L)$ tokenized with a smaller vocabulary $\mathcal{V}'$ (i.e., $|\mathcal{V}'|\ll|\mathcal{V}|$).
This has been shown effective for training sparse word-level ASR~\cite{higuchi2022hierarchical}.
The BERT-CTC loss is combined with the hierarchical CTC loss as
\begin{align}
    (1 - \lambda_{\mathsf{ctc}}) \mathcal{L}_{\mathsf{bc}} (O,W) + \lambda_{\mathsf{ctc}}\mathcal{L}_{\mathsf{ctc}}(O,W'), \label{eq:L_bertctc_hier}
\end{align}
where $\lambda_{\mathsf{ctc}}$ is a tunable parameter.
We investigate the importance of the hierarchical loss in~\cref{ssec:exp_ablation}.

\subsection{Inference}\label{ssec:bertctc_inference}
The most probable token sequence $\hat{W}$ is estimated by solving Eq.~\eqref{eq:asr} for Eq.~\eqref{eq:p_bertctc_W_O} as
\begin{align}
    \hat{W} &= \argmax_{W} \sum_{\tilde{W}} p(W|\tilde{W},O) p(\tilde{W}|O) \label{eq:bertctc_inference} \\
    &\approx \argmax_{W} p(W|\bar{W},O), \label{eq:bertctc_inference_approx} \\
    \text{where}&\ \ \bar{W} = \argmax_{\tilde{W}} p(\tilde{W}|O). \label{eq:bar_W}
\end{align}
From Eq.~\eqref{eq:bertctc_inference} to Eq.~\eqref{eq:bertctc_inference_approx},
we make the Viterbi approximation to deal with the intractable summation over all possible masked sequences.

To solve Eq.~\eqref{eq:bertctc_inference_approx},
we design a mask-predict algorithm~\cite{ghazvininejad2019mask} assisted by CTC inference,
inspired by~\cite{chan2020imputer,higuchi2020mask}.
See Table~\ref{tb:decoding_example} for an example decoding and Appendix~\ref{apdx:inference_algorithm} for pseudocode.
The algorithm first initializes a target sequence with an estimated length,
which is then followed by $k=\{1,\cdots,K\}$ iterations of token masking and prediction steps.

\paragraph{Initialization {\rm ($k=1$)}}
BERT-CTC is non-auto-regressive, and
the length of a target sequence $\hat{N}$ needs to be given in advance to start decoding~\cite{gu2018non}.
We determine the target length based on the auxiliary sequence $\hat{W}'$ predicted from the audio encoder output $H^{\mathsf{ae}}$ as $\hat{N}\sim|\hat{W}'|$.
Given the estimated length,
we initialize an initial masked sequence $\bar{W}^{(k=1)}$
by filling all $\hat{N}$ positions
with the mask token $\texttt{[MASK]}$.
By feeding $H^{\mathsf{ae}}$ and $H^{\mathsf{bert}}$ ($=\text{BERT}(\bar{W}^{(k=1)})$) to the self-attention module,
a hypothesized sequence $\hat{W}^{(k=1)}$ is obtained via CTC inference.
Here, $\hat{W}^{(k=1)}$ is predicted only from speech
without any observations from output tokens,
as they are all masked.

\paragraph{Token Masking Step {\rm (Eq.~\eqref{eq:bar_W})}}
Given a current prediction $\hat{W}^{(k)}$,
we replace $m(k)$ tokens having the lowest probability scores with $\texttt{[MASK]}$,
which results in the next masked sequence $\bar{W}^{(k+1)}$.
Here, $m(k)$ is a linear decay function $m(k)=\lfloor|\hat{W}^{(k)}|\cdot\frac{K-k}{K}\rfloor$, similar to~\cite{ghazvininejad2019mask}.

\vspace{-0.1cm}
\paragraph{Token Prediction Step {\rm (Eq.~\eqref{eq:bertctc_inference_approx})}}
$H^{\mathsf{ae}}$ and $H^{\mathsf{bert}}$ ($=\text{BERT}(\bar{W}^{(k+1)})$)
are fed to the self-attention module to generate the next hypothesis $\hat{W}^{(k+1)}$.
Here, the prediction of $\hat{W}^{(k+1)}$ is conditioned on the contextual embedding obtained from BERT.

\vspace{0.1cm}
Similar to~\cite{chan2020imputer,chi2021align},
BERT-CTC inference repeatedly predicts a target sequence at the alignment level,
which does not require an additional mechanism~\cite{gu2019levenshtein,higuchi2021improved} for adjusting the target length over iterations.
Moreover, BERT-CTC considers the output dependencies at the token level,
making it more suitable for a model to capture linguistic information.

\subsection{BERT-CTC for End-to-End SLU}
In addition to E2E-ASR,
BERT-CTC can model end-to-end SLU jointly by extending Eq.~\eqref{eq:p_bertctc_a_tW_O} as
\begin{align}
    \hspace{-0.05cm}
    p(y&|\text{BERT}(\tilde{W}),O) \nonumber \\
    &= \text{Softmax}(\text{SelfAttn}_{T+1}(H^{\mathsf{ae}}, H^{\mathsf{bert}})),
    \label{eq:p_y_W_O}
\end{align}
where we assume $y\in\mathcal{Y}$ as an intent label in a set of intents $\mathcal{Y}$.
Note that $\text{SelfAttn}_{T+1}(\cdot)$ indicates the $T+1$-th output of the self-attention module,
which corresponds to the \texttt{[CLS]} token of BERT.

\paragraph{Training}
The loss is defined by adding Eq.~\eqref{eq:L_bertctc_hier} and the negative log-likelihood of Eq.~\eqref{eq:p_y_W_O} as
\begin{align}
    \text{Eq.~\eqref{eq:L_bertctc_hier}} - \lambda_{\mathsf{slu}} \log p(y|\text{BERT}(\tilde{W}),O), \label{eq:L_bertctc_slu}
\end{align}
where $\lambda_{\mathsf{slu}}$ is a tunable parameter.

\paragraph{Inference}
The most probable label $\hat{y}$ can be estimated at any timing of BERT-CTC inference
by $\hat{y}=\argmax_{y} p(y|\bar{W},O)$.
When $k=1$, the label is predicted only from audio information, and
when $k=K$, the label is predicted with full access to audio and linguistic information.

\section{Additional Related Work}
\label{sec:related_work}
\paragraph{End-to-End ASR with MLM}
Inspired by the great success in non-autoregressive neural machine translation,
conditional masked language model (CMLM)~\cite{ghazvininejad2019mask} has been adopted for E2E-ASR.
Audio-CMLM (A-CMLM)~\cite{chen2020non} has trained an E2E-ASR model with an MLM objective~\cite{devlin2019bert},
making token predictions conditioned on both the speech input and a partially masked target sequence.
Imputer~\cite{chan2020imputer} and Mask-CTC~\cite{higuchi2020mask, higuchi2021improved} have introduced CTC to the CMLM-based modeling,
where the mask-predict algorithm is used to refine a frame-level or token-level sequence predicted by  CTC.

Our method of combining CTC and MLM is related to the above studies,
but conceptually different in that BERT-CTC aims to relax the conditional independence assumption used in CTC by leveraging an external pre-trained MLM (i.e., BERT) as contextual embedding.

\paragraph{LM Integration for End-to-End ASR.}
There is a line of prior studies seeking to integrate an external LM into E2E-ASR.
Shallow fusion has been the most widely used approach~\cite{hannun2014deep,gulcehre2015using,chorowski2017towards,kannan2018analysis},
which linearly interpolates the output probabilities from an E2E-ASR model and external LM.
Deep fusion~\cite{gulcehre2015using} is a more structured approach,
where an E2E-ASR model is jointly trained with an external LM to learn the optimal combination of the audio and linguistic information in a latent space.
Cold fusion~\cite{sriram2018cold} and component fusion~\cite{shan2019component} have further improved deep fusion by
a gating mechanism that learns a more sophisticated combination of the two models.

Our approach can be seen as a variant of cold fusion in that an external pre-trained MLM is fused to a CTC-based E2E-ASR model,
selectively combining audio and linguistic representations 
via the self-attention mechanism.
However,
BERT-CTC is a novel direction in which we seek to integrate BERT into a CTC-based model in a theoretically-sound manner.

\section{Experiments}
We used the ESPnet toolkit~\cite{watanabe2018espnet} for all the experiments.
All the implementations and recipes are made publicly available (see~\cref{sec:intro}).

\subsection{Tasks and Datasets}
\paragraph{Speech Recognition}
We evaluated models on the
LibriSpeech~\cite{panayotov2015librispeech}, TED-LIUM2~\cite{rousseau2014enhancing} and AISHELL-1~\cite{bu2017aishell} datasets.
LibriSpeech consists of read English speech from audiobooks, and
we used \textit{train-clean-100} for training.
TED-LIUM2 contains spontaneous English speech from Ted Talks.
AISHELL-1 consists of read Mandarin speech.

\vspace{-0.2cm}
\paragraph{Spoken Language Understanding}
We also evaluated our model on the SLURP dataset~\cite{bastianelli2020slurp}.
SLURP consists of English prompts of an in-home personal robot assistant, and
we focused on the intent classification task.

\vspace{0.1cm}
We used the standard development and test sets
for tuning hyper-parameters and evaluating performance for each dataset.
Full dataset descriptions are in Appendix~\ref{apdx:dataset}.

\subsection{End-to-End ASR Models}

\noindent\textbf{CTC (baseline)}:
A model trained based on the CTC loss $\mathcal{L}_{\mathsf{ctc}}$ (see~\cref{ssec:ctc}).
Given the recent advances in CTC-based modeling~\cite{higuchi2021comparative},
we built a strong baseline using the intermediate CTC technique~\cite{tjandra2020deja,lee2021intermediate},
which applies an auxiliary CTC loss to intermediate outputs of the audio encoder.
We used the intermediate loss in a hierarchical manner~\cite{sanabria2018hierarchical},
where the loss is calculated using a target sequence tokenized with a smaller vocabulary (i.e., $\mathcal{V}'$ in \cref{ssec:bertctc_training}).

\noindent\textbf{RNN-T (baseline)}:
A model trained based on the RNN-T loss $\mathcal{L}_{\mathsf{rnnt}}$ (see~\cref{ssec:rnnt}).
Considering the recent techniques developed upon multi-task learning~\cite{boyer2021study},
we trained a strong model using an auxiliary CTC loss applied to the audio encoder output~\cite{jeon2021multitask}.
Same as CTC, we enhanced the audio encoder with intermediate CTC~\cite{lee2022memory}.
All the CTC losses were calculated using the smaller-vocabulary sequence.

\noindent\textbf{BERT-CTC (ours)}:
The proposed model trained based on the BERT-CTC loss (Eq.~\eqref{eq:L_bertctc_hier}).
As in the other models, we adopted intermediate CTC for the audio encoder.
All the CTC losses were calculated using the smaller-vocabulary sequence.

See Appendices~\ref{apdx:interctc} and~\ref{apdx:model_details} for intermediate CTC and detailed model descriptions, respectively.

\begin{table*}[t]
    \centering
    \resizebox{.98\linewidth}{!}{
    \begin{tabular}{lcccccccc}
        \toprule
        \multirow{3}{*}[-6pt]{\textbf{Model}} & \multicolumn{4}{c}{\textbf{LibriSpeech-100h}} & \multicolumn{2}{c}{\textbf{TED-LIUM2}} & \multicolumn{2}{c}{\textbf{AISHELL-1}} \\
        \cmidrule(l{0.3em}r{0.3em}){2-5}\cmidrule(l{0.3em}r{0.3em}){6-7}\cmidrule(l{0.3em}r{0.3em}){8-9}
        & \multicolumn{2}{c}{Dev WER ($\downarrow$)} & \multicolumn{2}{c}{Test WER ($\downarrow$)} & \multirow{2}{*}[-3pt]{Dev WER ($\downarrow$)} & \multirow{2}{*}[-3pt]{Test WER ($\downarrow$)} & \multirow{2}{*}[-3pt]{Dev CER ($\downarrow$)} & \multirow{2}{*}[-3pt]{Test CER ($\downarrow$)} \\
        \cmidrule(l{0.3em}r{0.3em}){2-3}\cmidrule(l{0.3em}r{0.3em}){4-5}
        & clean & other & clean & other \\
        \midrule
        CTC$^\dagger$ & 11.2 & 21.4 & 11.4 & 22.0 & \pz9.9 & \pz9.3 & 5.1 & 5.6 \\
        RNN-T$^\dagger$ &\pz9.7 & 21.5 & \pz9.8 & 22.2 & 10.2 & \pz9.6 & 5.2 & 5.5 \\
        \midrule
        BERT-CTC & \textbf{\pz7.0} & \textbf{16.3} & \textbf{\pz7.2} & \textbf{16.6} & \textbf{\pz8.1} & \textbf{\pz7.6} & \textbf{3.9} & \textbf{3.9} \\
        \bottomrule
    \end{tabular}}
    \caption{WER [\%] on LibriSpeech-100h and TED-LIUM2, and CER [\%] on AISHELL-1. $\dagger$ indicates that the models are slightly different from the original CTC or RNN-T in that they are trained with hierarchical CTC loss.}
    \label{tb:main_results}
\end{table*}
\subsection{Experimental Settings}

\paragraph{Model Configuration}
For the audio encoder,
we adopted the Conformer architecture~\cite{gulati2020conformer},
which consisted of $12$ encoder blocks.
The prediction network in RNN-T was a single long short-term memory (LSTM) layer.
The self-attention module in BERT-CTC had $6$ Transformer encoder blocks, and
we used a BERT$_{\text{BASE}}$ model provided by HuggingFace~\cite{transformers2020wolf}.

\vspace{-0.15cm}
\paragraph{Tokenization}
For each language,
we used the same vocabulary as BERT for tokenizing target texts.
We also constructed a smaller-sized vocabulary $\mathcal{V}'$ for the hierarchical losses,
which is obtained by applying the byte pair encoding-based algorithm~\cite{sennrich2016neural} to the transcription of each dataset.

\vspace{-0.15cm}
\paragraph{Training}
We mostly followed ESPnet recipes provided for each dataset.
For BERT-CTC,
we set $\lambda_{\mathsf{ctc}}$ (in Eq.~\eqref{eq:L_bertctc_hier}) to $0.3$ for all the ASR tasks and
$\lambda_{\mathsf{slu}}$ (in Eq.~\eqref{eq:L_bertctc_slu}) to $1.0$ for the SLU task.

\vspace{-0.15cm}
\paragraph{Inference}
For CTC,
we performed the best path decoding (\cref{ssec:ctc}).
For RNN-T, we used the beam search decoding (\cref{ssec:rnnt}) with a beam size of $20$.
For BERT-CTC,
unless otherwise indicated,
the number of iterations $K$ was always set to $20$ (\cref{ssec:bertctc_inference}).

\vspace{0.1cm}
Detailed experimental settings for reproducibility are in Appendix~\ref{apdx:experimental_details}.

\section{Results}
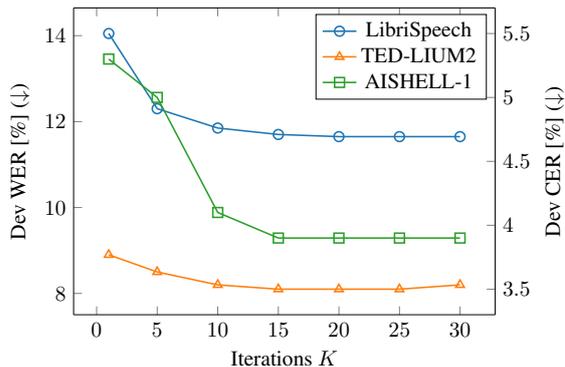
\begin{figure}[t]
\centering
\resizebox{0.99\columnwidth}{!}{
\begin{tikzpicture}
\definecolor{clr1}{RGB}{31, 119, 180}
\definecolor{clr2}{RGB}{255, 127, 14}
\definecolor{clr3}{RGB}{44, 160, 44}

\pgfplotsset{
  scale only axis,
}

\begin{axis}[
    height=5.5cm,
    axis y line*=left,
    xlabel=Iterations $K$,
    ylabel=Dev WER $\text{[\%]}$ ($\downarrow$),
]

\addplot[mark=o, mark size=2.5pt, clr1, thick]
  coordinates{
    (1, 14.05)
    (5, 12.3)
    (10, 11.85)
    (15, 11.7)
    (20, 11.65)
    (25, 11.65)
    (30, 11.65)
  };
\label{plot_1_y1}

\addplot[mark=triangle, mark size=2.5pt, clr2, thick]
  coordinates{
    (1, 8.9)
    (5, 8.5)
    (10, 8.2)
    (15, 8.1)
    (20, 8.1)
    (25, 8.1)
    (30, 8.2)
  }; \label{plot_2_y1}

\end{axis}

\begin{axis}[
    height=5.5cm,
    axis y line*=right,
    axis x line=none,
    ylabel=Dev CER $\text{[\%]}$ ($\downarrow$),
    ymin=3.3,
    ymax=5.7,
]
\addlegendimage{/pgfplots/refstyle=plot_1_y1}\addlegendentry{LibriSpeech}
\addlegendimage{/pgfplots/refstyle=plot_2_y1}\addlegendentry{TED-LIUM2}

\addplot[mark=square, mark size=2.5pt, clr3, thick]
  coordinates{
    (1, 5.3)
    (5, 5.0)
    (10, 4.1)
    (15, 3.9)
    (20, 3.9)
    (25, 3.9)
    (30, 3.9)
  }; \label{plot_1_y2}

\addlegendimage{/pgfplots/refstyle=plot_1_y2}\addlegendentry{AISHELL-1}

\end{axis}

\end{tikzpicture}
}
\vspace{-0.6cm}
\caption{BERT-CTC results on development sets, using different number of decoding iterations.}
\vspace{-0.1cm}
\label{fig:iterations}
\end{figure}

\paragraph{Speech Recognition}
Table~\ref{tb:main_results} shows results on
LibriSpeech-100h and TED-LIUM2 in word error rate (WER), and
AISHELL-1 in character error rate (CER).
While RNN-T slightly outperformed CTC on several evaluation sets in LibriSpeech-100h and AISHELL-1,
CTC resulted in better performance on TED-LIUM2.
RNN-T was ineffective at training ASR with the BERT vocabulary,
particularly when a severe mismatch exists against the target ASR domain (i.e., Wikipedia vs.\ lecture).
BERT-CTC significantly outperformed the baselines,
consistently achieving the best results on all datasets.
BERT-CTC improved over RNN-T, and
we attribute this to not only considering the whole context of the target sequence
but also using the powerful representations from BERT,
which we further analyze later.
In Appendix~\ref{apdx:comparison},
we compare our AISHELL-1 results to those from recent works and
show that our approach is on par with the state-of-the-art~\cite{zheng2021wav} with fewer parameters.
Figure~\ref{fig:iterations} illustrates the correlation between BERT-CTC results and the number of decoding iterations.
When decoded with $K=1$,
the model only uses speech input to predict a token sequence.
By increasing $K$,
the model beneficially exploited the BERT knowledge
for refining the output tokens.

\begin{table}[t]
    \centering
    \scalebox{0.82}{
    \begin{tabular}{lcc}
        \toprule
        \textbf{Model} & \textbf{WER} ($\downarrow$) & \textbf{Acc.} ($\uparrow$) \\
        \midrule
        ESPnet-SLU~\cite{arora2022espnet} & -- & 86.3 \\
        ASR + BERT~\cite{arora2022espnet} & -- & 85.7 \\
        \midrule
        BERT-CTC ($K=1$) & 19.1 & 87.0 \\
        BERT-CTC ($K=20$) & \textbf{18.2} & \textbf{87.8} \\
        \bottomrule
    \end{tabular}}
    \caption{WER [\%] and classification accuracy [\%] on SLURP intent classification task.}
    \label{tb:slurp}
\end{table}
\begin{table}[t]
    \centering
    \scalebox{0.85}{
    \begin{tabular}{lcc}
        \toprule
        \textbf{Model} & \textbf{Dev WER} ($\downarrow$) & \textbf{Test WER} ($\downarrow$) \\
        \midrule
        CTC$^\dagger$ & \textbf{11.2} / \textbf{21.4} & \textbf{11.4} / \textbf{22.0} \\
        w/o hierarchical loss & 11.8 / 23.2 & 12.2 / 24.1 \\
        \midrule
        RNN-T$^\dagger$ & \textbf{\pz9.7} / \textbf{21.5} & \textbf{\pz9.8} / \textbf{22.2} \\
        w/o hierarchical loss & 11.4 / 24.6 & 11.5 / 25.8 \\
        \midrule
        BERT-CTC & \textbf{\pz7.0} / \textbf{16.3} & \textbf{\pz7.2} / \textbf{16.6} \\
        w/o hierarchical loss & \pz8.6 / 19.1 & \pz8.9 / 19.5 \\
        w/o BERT & \pz7.4 / 17.2 & \pz7.4 / 17.7 \\
        \bottomrule
    \end{tabular}}
    \caption{Ablation studies on LibriSpeech-100h.}
    \label{tb:ablation}
\end{table}
\begin{table*}[t]
    \centering
    \resizebox{.99\linewidth}{!}{
    \begin{tabular}{ll}
        \toprule
        $k\!=\!1$ & ... \hl{thou} \hl{a} \hl{gave} \hl{meet} \hl{any} \hl{one} \hl{afterter} \hl{these} \hl{hour} \hl{recite} \hl{aught} \hl{of} \hl{courtry} \hl{whether} \hl{he} \hl{be} \hl{ne'er} ... \\
        $k\!=\!10$ & ... thou \hl{a} \hl{again} \hl{meet} \hl{any} \hl{one} \hl{afterter} \hl{these} \hl{hour} rec\hl{iteiting} \hl{aught} \hl{of} \hl{poetryry} \hl{whether} \hl{he} \hl{be} \hl{near'er} ...\\
        $k\!=\!15$ & ... thou \hl{again} meet any one after \hl{this} hour rec\hl{iteiting} aught of \hl{poetryry} whether he be \hl{near'or} ... \\
        $k\!=\!20$ & ... thou \blue{again} meet any one \blue{after} \blue{this} hour \red{reciteiting} aught of \blue{poetry} whether he be \blue{near} ... \\
        {\small w/o }BERT & ... thou \red{a} \red{gag} meet any one after this hour \red{residing} aught of \red{boy} whether he be near ... \\
        \cdashlinelr{1-2}
        Reference & ... thou again meet any one after this hour reciting aught of poetry whether he be near ... \\
        \bottomrule
    \end{tabular}}
    \caption{Decoding example from LibriSpeech test-other set (2033-164914-0016). At each iteration, the highlighted tokens are masked and repredicted in the next iteration. Blue indicates refined tokens, and red indicates ones not.}
    \label{tb:decoding_example}
\end{table*}
\paragraph{Spoken Language Understanding}
Table~\ref{tb:slurp} lists the results of the SLURP intent classification task, evaluated in accuracy.
We refer to the ESPnet-SLU~\cite{arora2022espnet} result as a baseline,
which performs SLU along with ASR by
prepending an intent label to the corresponding output sequence.
We also refer to the ESPnet-SLU result obtained by stacking BERT on top of an ASR model,
which was found to be less effective.
BERT-CTC outperformed the baselines by effectively incorporating acoustic and linguistic information.
By decoding in a single iteration ($K=1$),
BERT-CTC predicted an intent only from speech, and
the accuracy was already higher than those of baselines.
We observed a slight but clear gain by increasing $K$,
which improved both ASR and SLU performance thanks to BERT.
We note that our result outperforms the state-of-the-art 86.9\% reported in~\cite{seo2022integration}.

\vspace{-0.1cm}
\section{Analyses}

\subsection{Ablation Studies}
\label{ssec:exp_ablation}
To validate the effectiveness of our model design for BERT-CTC,
we conduct ablation studies (Table~\ref{tb:ablation}) on the usage of hierarchical loss and BERT.

\vspace{-0.05cm}
\paragraph{Hierarchical Loss}
We observed that hierarchical CTC helped all the models
improve their performance by a large margin.
As the vocabulary of BERT is generally too large for E2E-ASR,
the hierarchical modeling was crucial for predicting the sparse word-level tokens.
Moreover,
the result indicates that the hierarchical loss is effective for training an ASR model with a vocabulary from a different domain,
as there is a non-negligible domain-mismatch between the BERT training text and ASR transcription.

\vspace{-0.05cm}
\paragraph{BERT}
To ablate BERT-CTC with BERT,
we replaced $\text{BERT}(\cdot)$ in Eq.~\eqref{eq:H_bert} with a simple embedding layer with positional encoding.
We found that removing BERT led to degradation in BERT-CTC performance,
which supports the importance of using BERT.
However, interestingly, the result was still better than the baselines,
indicating the advantage over RNN-T in that BERT-CTC is capable of considering the bi-directional context.

\subsection{Error Analysis with Decoding Example}

Table~\ref{tb:decoding_example} shows a process of BERT-CTC inference,
decoding an utterance in the LibriSpeech test set.
In the output sequence at $k=1$,
the model mistakenly predicted phonetically similar tokens (e.g., ``again''$\rightarrow$``a gave'', ``near''$\rightarrow$``ne'er'').
At the first iteration, the model was only conditioned on acoustic information,
making it challenging to determine target tokens accurately.
As the iteration proceeded,
the model corrected the most errors by considering the output dependency.
Unlike the original mask-predict algorithm~\cite{ghazvininejad2019mask},
our approach permits for flexibly adjusting the target length,
enabling the model to resolve insertion and deletion errors
(e.g., ``afterer''$\rightarrow$``after'').
We also show an example obtained w/o BERT (from Table~\ref{tb:ablation}),
which failed to recover tokens that were correctly recognized by BERT-CTC with BERT.

\begin{figure}[t]
    \centering
    \includegraphics[width=0.99\linewidth]{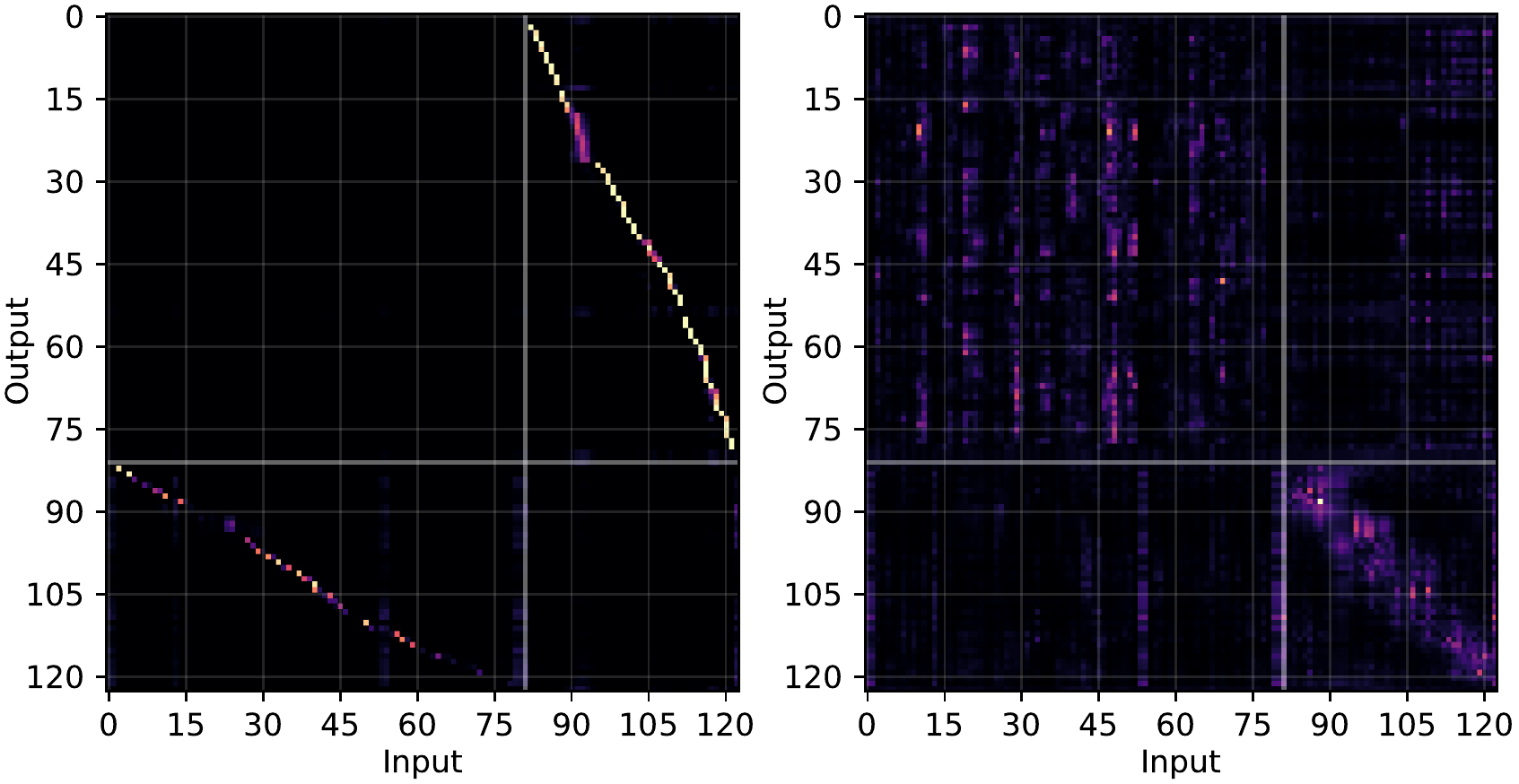}
    \vspace{-0.5cm}
    \caption{\mbox{Attention visualization for BERT-CTC. White} lines indicate the boundaries of audio and token seqs.}
    \label{fig:attention}
\end{figure}
\subsection{Conditional Independence of $p(W|\tilde{W}, \cancel{O})$}
\label{ssec:exp_cia_O}
We empirically validate the conditional independence assumption made in Eq.~\eqref{eq:p_W_tW_O_approx},
where the output sequence $W$ depends only on its masked sequence $\tilde{W}$ without audio information $O$.
To this end,
we augmented the BERT module by inserting adaptive cross-attention layers, which is
similar to Adapter-BERT Networks~\cite{guo2020incorporating}.
These additional layers are trained to infuse the audio encoder output $H^{\mathsf{ae}}$ into each BERT layer,
thereby allowing BERT-CTC to realize $p(W|\tilde{W},O)$.
When evaluated on LibriSpeech,
the modified BERT-CTC resulted in 7.2\%/17.9\% on the dev.\ set and 7.3\%/18.0\% on the test set,
which are worse than the results in Table~\ref{tb:main_results}.
This indicates that BERT already captures sophisticated linguistic information and
does not require extra parameters for adapting BERT to audio input.

\subsection{Attention Visualization}
Figure~\ref{fig:attention} depicts example attention weight matrices,
produced by the second self-attention layer of BERT-CTC.
We observed two major attention patterns:
weights aligning audio and token sequences by capturing inter-dependencies (Fig.~\ref{fig:attention} left) and
weights attending inner-dependencies within each sequence (Fig.~\ref{fig:attention} right).
These patterns support our motivation for the BERT-CTC design in learning inner/inter dependencies within/between the audio and token representations.

\subsection{Inference Speed Comparison}
To see how the iterative decoding with BERT affects the inference speed of BERT-CTC,
we evaluated each model on the real-time factor (RTF).
RTF was measured on the LibriSpeech test-other set using a single GPU with a batchsize of 1 or a single CPU.
RTFs for GPU / CPU inference resulted in
7.91e-3 / 4.18e-2 for CTC,
4.81e-1 / 4.55 for RNN-T, and
9.72e-2 / 7.22e-1 for BERT-CTC.
The semi-autoregressive characteristic in BERT-CTC enabled faster inference than autoregressive RNN-T and
provided further speedup with the parallel computing using GPU.

\section{Conclusion}
We proposed BERT-CTC that leverages BERT
for relaxing the conditional independence assumption in CTC.
BERT-CTC uses BERT as contextual embedding to explicitly condition CTC training/inference
on linguistic information.
Experimental results showed that
BERT-CTC improved over conventional approaches.
Moreover, we confirmed that BERT-CTC is applicable to end-to-end SLU.

\section*{Limitations}

\begin{table}[t]
    \centering
    \scalebox{0.82}{
    \begin{tabular}{llcc}
        \toprule
        & & \textbf{Dev WER}\ ($\downarrow$) & \textbf{Test WER} ($\downarrow$) \\
        $\mathcal{V}$ & \textbf{Model} & clean / other & clean / other \\
        \midrule
        $\mathcal{V}^{\mathsf{asr}}$ & CTC$^\dagger$ & \pz6.9 / 20.1 & \pz7.0 / 20.2 \\
        $\mathcal{V}^{\mathsf{asr}}$ & RNN-T$^\dagger$ & \textbf{\pz5.7} / 17.0 & \textbf{\pz6.0} / 17.2 \\
        $\mathcal{V}^{\mathsf{bert}}$ & CTC$^\dagger$ & 11.2 / 21.4 & 11.4 / 22.0 \\
        $\mathcal{V}^{\mathsf{bert}}$ & RNN-T$^\dagger$ & \pz9.7 / 21.5 & \pz9.8 / 22.3 \\
        \midrule
        $\mathcal{V}^{\mathsf{bert}}$ & BERT-CTC & \pz7.0 / \textbf{16.3} & \pz7.2 / \textbf{16.6} \\
        \bottomrule
    \end{tabular}}
    \caption{WER [\%] on LibriSpeech-100h. $\mathcal{V}^{\mathsf{asr}}$ indicates a subword vocabulary constructed from ASR transcriptions, where $|\mathcal{V}^{\mathsf{asr}}|=300$. $\mathcal{V}^{\mathsf{bert}}$ indicates the BERT vocabulary, where $|\mathcal{V}^{\mathsf{bert}}|=30522$.}
    \label{tb:asr_vocab}
\end{table}
\paragraph{Vocabulary Constraint}
The output unit of BERT-CTC is constrained to the vocabulary of BERT,
which is likely to be not generalized to an ASR domain and too sparse for ASR training.
Table~\ref{tb:asr_vocab} shows results on LibriSpeech-100h with different vocabularies,
where $\mathcal{V}^{\mathsf{asr}}$ is an ASR vocabulary with a vocabulary size of 300 constructed from LibriSpeech transcriptions, and
$\mathcal{V}^{\mathsf{bert}}$ is the BERT vocabulary with a vocabulary size of 30522.
We observed that, by using $\mathcal{V}^{\mathsf{asr}}$,
the performance of CTC and RNN-T improved over the results using $\mathcal{V}^{\mathsf{bert}}$ and closed the gap with the BERT-CTC results.
We believe that using a BERT variant with a smaller vocabulary, e.g., CharacterBERT~\cite{boukkouri2020character} improves BERT-CTC further.

\paragraph{Computational Cost}
BERT-CTC requires a high computational cost,
especially during inference,
due to the iterative forward calculations of BERT (i.e., $K\!=\!20$ times)
with the $\mathcal{O}(N^2)$ computational and memory complexities in the self-attention layers.
Still, GPUs can greatly accelerate the inference speed, and
BERT-CTC can alternatively use other pre-trained MLMs with lighter weights,
e.g., ALBERT~\cite{lan2019albert} and DistilBERT~\cite{sanh2019distilbert}.

\paragraph{Non-streaming}
BERT-CTC is not suited for online streaming scenarios,
where output tokens are predicted synchronously to sequential speech input.
It is not a significant problem when we consider applying BERT-CTC to utterance-level ASR tasks,
such as end-to-end SLU as we demonstrated the capability of BERT-CTC (Table~\ref{tb:slurp}).
Otherwise,
we can adopt existing techniques for making BERT-CTC streaming,
e.g., causal masking~\cite{vaswani2017attention}, time-restricted attention~\cite{povey2018time}, and
block-wise processing~\cite{tsunoo2019transformer}.
Another solution can be to apply the two-pass algorithm~\cite{sainath2019two},
where BERT-CTC first performs streaming recognition at $k=1$ and then refines the outputs using the full context information at $k>1$.

\section*{Acknowledgements}
This work was supported in part by JST ACT-X (JPMJAX210J) and JSPS KAKENHI (JP21J23495).
This work used the Extreme Science and Engineering Discovery Environment (XSEDE) \cite{xsede} supported by National Science Foundation grant number ACI-1548562. It uses the Bridges system \cite{nystrom2015bridges} supported by NSF award number ACI-1445606, at the Pittsburgh Supercomputing Center (PSC).
The authors would like to thank Jing Liu for providing insightful comments on the BERT-CTC formulation.

\bibliography{anthology,custom}
\bibliographystyle{acl_natbib}

\vfill
\appendix

\section{Inference Algorithm}
\label{apdx:inference_algorithm}
\begin{algorithm*}[t]
    \renewcommand{\ttdefault}{cmtt}
    \caption{\bf BERT-CTC Inference}
    \label{algo:bertctc_inference}
    \begin{algorithmic}[1]
    \renewcommand{\algorithmicrequire}{\textbf{Input:}}
    \renewcommand{\algorithmicensure}{\textbf{Output:}}
        \Statex \textbf{Input:} The number of iterations $K$, audio encoder output $H^{\mathsf{ae}}$
        \State $\hat{A}'=\argmax_{A'}p(A'|O)$ \Comment{Obtain the most probable alignment from the audio encoder}
        \State $\hat{W}'=\mathcal{B}_{\mathsf{ctc}}(\hat{A}')$
        \State $\hat{N} \sim |\hat{W}'|$ \Comment{Obtain the target length from the intermediate prediction}
        \State $\bar{W} = (w_n\!=\!\texttt{[MASK]} | n=1,\cdots,\hat{N})$ \quad \Comment{Initialize a masked sequence}
        \For {$k=1,\dots,K$}
        \Statex \vspace{-10pt}
        \State \blue{\texttt{\# Token prediction}}
        \State $H^{\mathsf{bert}} = \text{BERT}(\bar{W})$ \Comment{Forward BERT}
        \State $p(A|\cdot) = \text{Softmax}(\text{SelfAttn}(H^{\mathsf{ae}}, H^{\mathsf{bert}}))$ \quad \Comment{Forward self-attention module}
        \State $\hat{A} = \argmax_{A}p(A|\cdot)$ \Comment{Obtain the most probable alignment}
        \State $\hat{W}=\mathcal{B}_{\mathsf{ctc}}(\hat{A})$
        \Statex \vspace{-10pt}
        \State \blue{\texttt{\# Token-level probability calculation}}
        \State $\hat{P} = (\hat{p}_n = 0|n=1,\cdots,|\hat{W}|)$ \quad \Comment{Initialize \textit{token-level} probabilities}
        \State $n = 1$ \Comment{Initialize an index for token position}
        \State $a_0 = \epsilon$
        \For {$t=1,\dots,T$}
            \If{$\hat{a}_t = \epsilon$}
                \If{$\hat{a}_{t-1} \ne \epsilon$}
                    \State $n = n + 1$
                \EndIf
            \Else
                \State $\hat{p}_n = \mathrm{max}(p(a_t=\hat{w}_n|\cdot), \hat{p}_n)$ \quad \Comment{Keep the maximum probability for each token}
            \EndIf
        \EndFor
        \Statex \vspace{-10pt}
        \State \blue{\texttt{\# Token masking}}
        \State $M = \lfloor |\hat{W}| \cdot \frac{K - k}{K} \rfloor$ \Comment{Calculate the number of masked tokens}
        \State $\bar{W}=\text{MaskLowestProb}(\hat{W}, \hat{P}, M)$ \quad \Comment{Mask tokens with the $M$ lowest probability scores}
        \EndFor \\
        \Return $\hat{W}$
    \end{algorithmic}
\end{algorithm*}
Algorithm~\ref{algo:bertctc_inference}
describes the overall process of BERT-CTC inference.
For estimating the target length in line 3, at the implementation level,
we first decode $\hat{W}'$ into a sentence,
which is then tokenized using the BERT vocabulary, and
the length of the resulting sequence is used as the target length.
In lines 12--25, before the token masking step,
we calculate a probability score $\hat{p}_n$ for each token $\hat{w}_n$ in the estimated output sequence $\hat{W}$.
This score calculation simply takes the maximum value in frame-level token probabilities 
that correspond to a predicted token $\hat{w}_n$ after the collapsing operation.
In line 28,
given the probability scores,
$\text{MaskLowestProb}(\cdot)$ masks tokens in $\hat{W}$ with the $M$ lowest scores.

\section{Intermediate CTC}
\label{apdx:interctc}
Intermediate CTC~\cite{tjandra2020deja,lee2021intermediate} applies additional CTC losses to intermediate layers of the audio encoder network.
Let $H^{(e)}=(\bm{h}_t^{(e)}\in\mathbb{R}^{d^{\mathsf{ae}}}|t=1,\cdots,T)$ be an intermediate output of the $e$-th layer of the audio encoder,
which is computed as in Eq.~\eqref{eq:h_t} as
\begin{align}
    \bm{\mathrm{h}}_t^{(e)} \sim \text{AudioEnc}^{(e)}(O), \label{eq:h_t_e}
\end{align}
where $\text{AudioEnc}^{(e)}(\cdot)$ indicates the $e$-th layer output of the audio encoder.
Similar to Eq.~\eqref{eq:p_ctc_a_o},
token emission probabilities at each time frame is computed based on Eq.~\eqref{eq:h_t_e} as
\begin{align}
    p^{(e)}(a_t|O) &= \text{Softmax}(\text{Linear}(\bm{\mathrm{h}}^{(e)}_t)).
\end{align}
Finally, an intermediate CTC loss $\mathcal{L}_{\mathsf{ctc}}^{(e)}$ is defined similarly to Eq.~\eqref{eq:L_ctc} as
\begin{align}
    \mathcal{L}_{\mathsf{ctc}}^{(e)} (O, W) = -\log \sum_{A \in \mathcal{B}_{\mathsf{ctc}}^{-1}(W)} \prod_{t=1}^{T} p^{(e)}(a_t|O).
    \label{eq:L_interctc}
\end{align}

\section{Model Details}
\label{apdx:model_details}

\subsection{CTC (baseline)}
We applied the intermediate CTC loss to the 6-th layer of the audio encoder,
which is calculated using the smaller-vocabulary sequence $W'$ in a hierarchical multi-tasking manner.
Using the intermediate loss,
the CTC loss $\mathcal{L}_{\mathsf{ctc}}$ is extended as
\begin{align}
    (1 - \lambda_{\mathsf{ic}}) \mathcal{L}_{\mathsf{ctc}} (O,W) + \lambda_{\mathsf{ic}} \mathcal{L}^{(e=6)}_{\mathsf{ctc}} (O,W'), \label{eq:L_ctc_ic}
\end{align}
where $\lambda_{\mathsf{ic}}$ is a tunable weight for the intermediate loss, and we equally weighted each loss (i.e., $\lambda_{\mathsf{ic}} = 0.5$) as in~\cite{lee21layer,higuchi2022hierarchical}.

\subsection{RNN-T (baseline)}
We applied auxiliary CTC losses to the final and intermediate layer of the audio encoder.
As in Eq.~\eqref{eq:L_ctc_ic}, the intermediate loss was applied to the 6-th layer.
With the additional CTC losses,
the RNN-T loss $\mathcal{L}_{\mathsf{rnnt}}$ is extended as
\begin{align}
    &(1 - \lambda_{\mathsf{ctc}})\mathcal{L}_{\mathsf{rnnt}}(O,W) \nonumber \\
    &+\!\lambda_{\mathsf{ctc}} \{ (1\!-\!\lambda_{\mathsf{ic}}) \mathcal{L}_{\mathsf{ic}} (O,W')
    \!+\!\lambda_{\mathsf{ic}} \mathcal{L}^{(e=6)}_{\mathsf{ctc}} (O,W') \},
\end{align}
where $\lambda_{\mathsf{ctc}}$ is a tunable weight for CTC losses, and we set $\lambda_{\mathsf{ctc}} = 0.3$ as in~\cite{boyer2021study}.
Note that all the CTC losses were calculated using the smaller-vocabulary sequence $W'$ in a hierarchical multi-tasking manner.

\subsection{BERT-CTC (ours)}
We applied auxiliary CTC losses to the final and intermediate layer of the audio encoder.
As in Eq.~\eqref{eq:L_ctc_ic}, the intermediate loss was applied to the 6-th layer.
With the additional CTC losses,
the BERT-CTC loss $\mathcal{L}_{\mathsf{bc}}$ is extended as
\begin{align}
    &(1 - \lambda_{\mathsf{ctc}}) \mathcal{L}_{\mathsf{bc}} (O,W) \nonumber \\
    &+\!\lambda_{\mathsf{ctc}} \{ (1\!-\!\lambda_{\mathsf{ic}}) \mathcal{L}_{\mathsf{ic}} (O,W')
    \!+\!\lambda_{\mathsf{ic}} \mathcal{L}^{(e=6)}_{\mathsf{ctc}} (O,W') \},
\end{align}
where $\lambda_{\mathsf{ctc}}$ is a tunable weight for CTC losses, and we set $\lambda_{\mathsf{ctc}} = 0.3$.
Note that all the CTC losses were calculated using the smaller-vocabulary sequence $W'$ in a hierarchical multi-tasking manner (as explained in~\cref{ssec:bertctc_training}).

\section{Experimental Details}
\label{apdx:experimental_details}
\subsection{Dataset}
\label{apdx:dataset}
\begin{table*}[t]
    \centering
    \resizebox{.98\linewidth}{!}{
    \begin{tabular}{lccccc}
        \toprule
        \textbf{Dataset} & \textbf{Language} & \textbf{Speech Style} & \textbf{\# Train Hours} & \textbf{\# Valid. Hours} & \textbf{\# Test Hours} \\
        \midrule
        LibriSpeech-100h~\cite{panayotov2015librispeech} & English & Read & 100h & 5.4h / 5.3h & 5.4h / 5.1h \\
        TED-LIUM2~\cite{rousseau2014enhancing} & English & Spontaneous & 210h & 1.6h & 2.6h \\
        AISHELL-1~\cite{bu2017aishell} & Mandarin & Read & 170h & 10h & 5h \\
        \bottomrule
    \end{tabular}}
    \caption{ASR dataset descriptions. The validation and test sets of LibriSpeech are split into ``clean'' / ``other'' sets based on the quality of the recordec utterances.}
    \label{tb:asr_dataset}
\end{table*}
\begin{table*}[t]
    \centering
    \resizebox{.98\linewidth}{!}{
    \begin{tabular}{lccccc}
        \toprule
        \textbf{Dataset} & \textbf{Language} & \textbf{\# Intents} & \textbf{\# Train Hours} & \textbf{\# Valid. Hours} & \textbf{\# Test Hours} \\
        \midrule
        SLURP~\cite{bastianelli2020slurp} & English & 69 & 40h + 43h & 6.9h & 10.3h \\
        \bottomrule
    \end{tabular}}
    \caption{Dataset description for SLURP intent classification. We bootstrap the train set with the synthetic data.}
    \label{tb:slu_dataset}
\end{table*}
Tables~\ref{tb:asr_dataset} and~\ref{tb:slu_dataset} list descriptions of ASR and SLU datasets, respectively.
Data preparation was done using the ESPnet2 recipe provided for each dataset:
LibriSpeech-100h\footnote{\url{https://github.com/espnet/espnet/tree/master/egs2/librispeech_100/asr1}},
TED-LIUM2\footnote{\url{https://github.com/espnet/espnet/tree/master/egs2/tedlium2/asr1}},
AISHELL-1\footnote{\url{https://github.com/espnet/espnet/tree/master/egs2/aishell/asr1}},
SLURP\footnote{\url{https://github.com/espnet/espnet/tree/master/egs2/slurp/asr1}}.

\subsection{Model Configuration}
\label{apdx:model_conf}
For the audio encoder network,
we used the Conformer~\cite{gulati2020conformer}-based encoder architecture implemented in ESPnet~\cite{guo2021recent}.
The audio encoder consisted of $2$ or $3$ convolutional neural network (CNN) layers followed by a stack of $12$ encoder blocks.
The dimensions of the self-attention layer $d^{\mathsf{ae}}$ and
feed-forward network $d^{\mathsf{ff}}$ were set to $256$ and $1024$, respectively, and
the number of heads $d^{\mathsf{head}}$ was set to $4$.
The kernel size of depthwise separable convolution was set to $31$.
For RNN-T,
the prediction network was a single LSTM layer with $512$ units, and 
the joint network was a single linear layer with $640$ units.
For BERT-CTC,
we built the Transformer~\cite{vaswani2017attention}-based architecture for the self-attention module,
which consisted of a stack of $6$ encoder blocks with $d^{\mathsf{model}}=256$, $d^{\mathsf{ff}}=2048$, and $d^{\mathsf{head}}=4$.
Before feeding into the self-attention module,
the hidden vectors, $H^{\mathsf{ae}}$ and $H^{\mathsf{bert}}$, were emebbeded using
$2$ CNN layers and a single linear layer, respectively,
which mapped each vector to the dimension size of $d^{\mathsf{model}}$.
For the BERT module in BERT-CTC,
we downloaded pre-trained models from the HuggingFace Transformers library~\cite{transformers2020wolf}\footnote{\url{https://github.com/huggingface/transformers}}.
We used a BERT$_{\text{BASE}}$ model provided for each language: English\footnote{\url{https://huggingface.co/bert-base-uncased}}, Mandarin\footnote{\url{https://huggingface.co/bert-base-chinese}}.
Note that the dimension of the BERT output $d^{\mathsf{bert}}$ was $768$.
The number of total/trainable parameters in the CTC, RNN-T, and BERT-CTC models was about $30$M/$30$M, $60$M/$60$M, and $150$M/$40$M, respectively.

\subsection{Tokenization}
We used the same subword vocabulary as BERT for tokenizing target texts,
where the vocabulary size $|\mathcal{V}|$ was $30522$ for English and $21128$ for Mandarin.
For the smaller-sized vocabulary $\mathcal{V}'$ used in hierarchical CTC,
we used SentencePiece~\cite{kudo2018subword}\footnote{\url{https://github.com/google/sentencepiece}} to construct subword vocabularies from transcription data in each training set.
Following the ESPnet recipes,
the vocabulary size was set to $300$ for LibriSpeech-100h, and
$500$ for TED-LIUM2 and SLURP.
For AISHELL-1,
we used character-level tokenization with $4231$ Chinese characters.

\subsection{Training}
\begin{table}[t]
    \centering
    \scalebox{0.8}{
    \begin{tabular}{lr}
        \toprule
        \textbf{Hyperparameter} & \textbf{Value} \\
        \midrule
        Dropout rate & 0.1 \\
        LR schedule & Noam~\cite{vaswani2017attention} \\
        Max learing rate & best of [1e-3, 2e-3] \\
        Warmup steps & 15k \\
        Epochs & best of [50, 70, 100]\\
        Adam betas & (0.9, 0.98)\\
        Weight decay & 1e-6 \\
        \bottomrule
    \end{tabular}}
    \caption{Training configuration for CTC model.}
    \label{tb:ctc_config}
\end{table}
\begin{table}[t]
    \centering
    \scalebox{0.8}{
    \begin{tabular}{lr}
        \toprule
        \textbf{Hyperparameter} & \textbf{Value} \\
        \midrule
        Dropout rate & 0.1 \\
        LR schedule & Noam~\cite{vaswani2017attention} \\
        Max learing rate & 2e-3 \\
        Warmup steps & 15k \\
        Epochs & best of [50, 70]\\
        Adam betas & (0.9, 0.98)\\
        Weight decay & 1e-6 \\
        \bottomrule
    \end{tabular}}
    \caption{Training configuration for RNN-T model.}
    \label{tb:rnnt_config}
\end{table}
\begin{table}[t]
    \centering
    \scalebox{0.8}{
    \begin{tabular}{lr}
        \toprule
        \textbf{Hyperparameter} & \textbf{Value} \\
        \midrule
        Dropout rate & 0.1 \\
        LR schedule & Noam~\cite{vaswani2017attention} \\
        Max learing rate & best of [1e-3, 2e-3] \\
        Warmup steps & 15k \\
        Epochs & best of [50, 70, 100]\\
        Adam betas & (0.9, 0.98)\\
        Weight decay & 1e-6 \\
        \bottomrule
    \end{tabular}}
    \caption{Training configuration for BERT-CTC model.}
    \label{tb:bertctc_config}
\end{table}
\begin{table*}[t]
    \centering
    \resizebox{.99\linewidth}{!}{
    \begin{tabular}{lccccccc}
        \toprule
        \multirow{2}{*}[-4pt]{\textbf{Model}} & \multirow{2}{*}[-4pt]{\shortstack{\textbf{\#params} [$\mathrm{M}$]\\Total (Trainable)}} & \multicolumn{2}{c}{\textbf{Pre-trained}} & \multirow{2}{*}[-4pt]{\textbf{Dev CER} ($\downarrow$)} & \multirow{2}{*}[-4pt]{\textbf{Test CER} ($\downarrow$)} \\
        \cmidrule(l{0.3em}r{0.3em}){3-4}
        & & AM & LM & & \\
        \midrule
        rePLM-NAR-ASR~\cite{yu2022non} & 120 (120) & -- & BERT & 4.2 & 4.8 \\
        CTC/Attention~\cite{deng2021improving} & 161 (152) & wav2vec2.0 & -- & 4.7 & 5.0 \\
        CTC/Attention~\cite{deng2021improving} & 218 (209) & wav2vec2.0 & DistilGPT2 & 3.9 & 4.2 \\
        NAR-CTC/Attention~\cite{deng2022improving} & 204 (195) & wav2vec2.0 & BERT & 4.0 & 4.3 \\
        Wav-BERT~\cite{zheng2021wav} & 380 (380) & wav2vec2.0 & BERT & \textbf{3.6} & \textbf{3.8} \\
        \midrule
        BERT-CTC (ours) & 143 (40) & -- & BERT & 3.9 & 3.9 \\
        \bottomrule
    \end{tabular}}
    \caption{Comparison to prior works on AISHELL-1. The number of trainable parameters in BERT-CTC is fewer than in the others because BERT-CTC uses BERT without fine-tuning.}
    \label{tb:aishell}
\end{table*}
\begin{table*}[t]
    \centering
    \resizebox{.99\linewidth}{!}{
    \begin{tabular}{lccccccc}
        \toprule
        \multirow{3}{*}[-6pt]{\textbf{Model}} & \multirow{3}{*}[-6pt]{\textbf{\#iters}} & \multicolumn{4}{c}{\textbf{LibriSpeech-100h}} & \multicolumn{2}{c}{\textbf{TED-LIUM2}} \\
        \cmidrule(l{0.3em}r{0.3em}){3-6}\cmidrule(l{0.3em}r{0.3em}){7-8}
        & & \multicolumn{2}{c}{Dev WER ($\downarrow$)} & \multicolumn{2}{c}{Test WER ($\downarrow$)} & \multirow{2}{*}[-3pt]{Dev WER ($\downarrow$)} & \multirow{2}{*}[-3pt]{Test WER ($\downarrow$)} \\
        \cmidrule(l{0.3em}r{0.3em}){3-4}\cmidrule(l{0.3em}r{0.3em}){5-6}
        & & clean & other & clean & other \\
        \midrule
        Mask-CTC~\cite{higuchi2020mask} & 10 & 7.2 & 20.3 & 7.5 & 20.6 & 8.9 & 8.5 \\
        Improved Mask-CTC~\cite{higuchi2021improved} & 5 & 7.0 & 19.8 & 7.3 & 20.2 & 8.8 & 8.3 \\
        Align-Denoise~\cite{chen2021align} & 1 & 8.0 & 22.3 & 8.4 & 22.5 & 9.0 & 8.7 \\
        Intermediate CTC~\cite{lee2021intermediate} & 1 & 6.9 & 19.7 & 7.1 & 20.2 & 8.5 & 8.3 \\
        Self-conditioned CTC~\cite{nozaki2021relaxing} & 1 & \textbf{6.6} & 19.4 & \textbf{6.9} & 19.7 & 8.7 & 8.0 \\
        KERMIT~\cite{fujita2020insertion} & $\simeq \log_2 (N)$ & 7.1 & 19.7 & 7.4 & 20.2 & 9.1 & 8.2 \\
        \midrule
        BERT-CTC (ours) & 20 & 7.0 & \textbf{16.3} & 7.2 & \textbf{16.6} & \textbf{8.1} & \textbf{7.6} \\
        \bottomrule
    \end{tabular}}
    \caption{Comparison of BERT-CTC and non-autoregressive E2E-ASR models on LibriSpeech-100h and TED-LIUM2. The prior results are obtained from the comparative study conducted in~\cite{higuchi2021comparative}.}
    \label{tb:nar}
\end{table*}
All the models were implemented and trained using ESPnet~\cite{watanabe2018espnet}\footnote{\url{https://github.com/espnet/espnet}} and PyTorch~\cite{paszke2019pytorch}\footnote{\url{https://github.com/pytorch/pytorch}}.
In Tables~\ref{tb:ctc_config},~\ref{tb:rnnt_config}, and~\ref{tb:bertctc_config},
we summarize training configurations for the CTC, RNN-T, and BERT-CTC models, respectively.
We augmented speech data using speed perturbation~\cite{ko2015audio} with a factor of $3$ and SpecAugment~\cite{park2019specaugment}.
For the hyperparameters in SpecAugment,
we set the number of frequency and time masks to $2$ and $5$, and
the size of frequency and time masks to $27$ and $0.05T$.
Note that the maximum size of the time mask depends on the utterance length $T$.
After training,
model parameters were averaged over 10 checkpoints with the best validation performance.
For CTC,
we trained models using a single RTX 2080 Ti GPU for 1 to 3 days, depending on the tasks and number of epochs.
For RNN-T,
we trained models using 4 V100 GPUs for 5 to 7 days, depending on the tasks and number of epochs.
For BERT-CTC,
we trained models using a single RTX 2080 Ti GPU for 3 to 5 days, depending on the tasks and number of epochs.

\subsection{Inference}
RTF was measured using a single V100 GPU (with a batchsize of 1) or a single Intel(R) Xeon(R) Gold 6148 CPU@2.4 GHz.

\section{Comparison to Prior Works}
\label{apdx:comparison}
\paragraph{AISHELL-1}
Table~\ref{tb:aishell} lists results on AISHELL-1,
comparing our BERT-CTC with recent approaches using a pre-trained acoustic model (AM) or/and LM.
BERT-CTC achieved comparable performance to the state-of-the-art approach, Wav-BERT~\cite{zheng2021wav},
without using a pre-trained AM.
Moreover,
the number of trainable parameters in BERT-CTC was much fewer than in the other models because BERT-CTC used BERT as contextual embedding (without fine-tuning).
We attribute this advantage of BERT-CTC to our well-defined formulation for conditioning CTC training/inference with BERT knowledge.

\paragraph{Non-autoregressive End-to-End ASR}
Table~\ref{tb:nar} compares our BERT-CTC with the previous non-autoregressive E2E-ASR models on LibriSpeech-100h and TED-LIUM2.
It should be noted that we refer to~\cite{higuchi2021comparative} for the prior results, and
the comparison is not necessarily in an equivalent setting, e.g., we conducted experiments using ESPnet2 while the previous work used ESPnet1.
Overall, BERT-CTC achieved better results than the other non-autoregressive models,
thanks to the usage of BERT.
In particular, we observed clear differences in the LibriSpeech ``other'' sets and TED-LIUM2.
However, the performance on the LibriSpeech ``clean'' set was on par with the other approaches,
which we attribute to the vocabulary mismatch problem we have discussed in the limitation section.

\end{document}